\documentclass[fleqn,10pt]{wlscirep}
\usepackage[utf8]{inputenc}
\usepackage[T1]{fontenc}

\usepackage{graphicx}
\usepackage{dcolumn}
\usepackage{bm}
\usepackage{epsfig}
\usepackage{color}
\usepackage{longtable}
\usepackage{amsmath}
\usepackage{multirow}
\usepackage{tabularx}
\DeclareMathAlphabet{\mathpzc}{OT1}{pzc}{m}{it}

\definecolor{aogreen}{rgb}{0.0, 0.5, 0.0}

\def\ketm#1{  \left\vert  #1   \right\rangle   }

\def\sprm#1#2{  \left\langle #1 \left\vert \right. #2 \right\rangle   }

\def\mem#1#2#3{  \left\langle #1 \left\vert  #2 \right\vert #3 \right\rangle   }
\def\memred#1#2#3{  \left\langle #1 \vert\vert  #2 \vert\vert #3 \right\rangle   }

\definecolor{mymainmessagecolor}{RGB}{10,200,10}
\DeclareMathOperator*{\SumInt}{%
\mathchoice%
  {\ooalign{$\displaystyle\sum$\cr\hidewidth$\displaystyle\int$\hidewidth\cr}}
  {\ooalign{\raisebox{.14\height}{\scalebox{.7}{$\textstyle\sum$}}\cr\hidewidth$\textstyle\int$\hidewidth\cr}}
  {\ooalign{\raisebox{.2\height}{\scalebox{.6}{$\scriptstyle\sum$}}\cr$\scriptstyle\int$\cr}}
  {\ooalign{\raisebox{.2\height}{\scalebox{.6}{$\scriptstyle\sum$}}\cr$\scriptstyle\int$\cr}}
}

\title{Breakdown of the electric dipole approximation at Cooper minima in direct two-photon ionisation}

\author[1,2,3,*]{J.~Hofbrucker}
\author[1,2]{A.~V.~Volotka}
\author[1,2,3]{S.~Fritzsche}

\affil[1]{Helmholtz-Institut Jena, Fr\"o{}belstieg 3, D-07743 Jena, Germany}
\affil[2]{GSI Helmholtzzentrum f\"ur Schwerionenforschung GmbH, Planckstrasse 1, D-64291 Darmstadt, Germany}
\affil[3]{Theoretisch-Physikalisches Institut, Friedrich-Schiller-Universit\"at Jena, Max-Wien-Platz 1, D-07743 Jena, Germany}

\affil[*]{jiri.hofbrucker@uni-jena.de}

\begin{abstract}
We predict breakdown of the electric dipole approximation at nonlinear Cooper minimum in direct two-photon $K-$shell atomic ionisation by circularly polarised light. According to predictions based on the electric dipole approximation, we expect that tuning the incident photon energy to the Cooper minimum in two-photon ionisation results in pure depletion of one spin projection of the initially bound $1s$ electrons, and hence, leaves the ionised atom in a fully oriented state. We show that by inclusion of electric quadrupole interaction, dramatic drop of orientation purity is obtained. The low degree of the remaining ion orientation provides a direct access to contributions of the electron-photon interaction beyond the electric dipole approximation in the two-photon ionisation of atoms and molecules. The orientation of the photoions  can be experimentally detected either directly by a Stern-Gerlach analyzer, or by means of subsequent $K\alpha$ fluorescence emission, which has the information about the ion orientation imprinted in the polarisation of the emitted photons.
\end{abstract}

\begin{document}
\flushbottom
\maketitle
%
%
\thispagestyle{empty}


\section*{Introduction}

The interest in the inner-shell dynamics of atoms and molecules has been rising ever since excitation and ionisation of the strongly bound electrons became accessible by the first XUV and x-ray light sources \cite{Schnopper:1963:2558, Carlson:1965:A1655}. Nowadays, it is also possible to probe these systems in nonlinear regime with free-electron lasers \cite{Mcneil:2010:814, Allaria:2013:913}. 
That is why, in the last decade, much of experimental efforts \cite{Sorokin:2007:213002,  Richter:2010:194005, Doumy:2011:083002, Fukuzawa:2013:173005, Ma:2013:164018, Zitnik:2014:193201,Tamasaku:2014:10.1038, Szlachetko:2016:33292, Ghimire:2016:043418, Tamasaku:2018:083901, Royce:2018:112} have been paid to studying the fundamental properties of nonlinear light-matter interaction, and finding use in applied fields such as nonlinear spectroscopy. The theoretical developments were in many respects following the experimental trail. Starting with the pioneering work of Zernik \cite{Zernik:1964:A51}, who calculated the nonresonant two-photon ionisation cross section for hydrogen, and continuing with further theoretical development into direction of many-photon absorption and ionisation of complex atoms. In particular, significant progress has been made in the description of the two-photon ionisation of outer-shell electrons of noble gas atoms, where electron correlations play a significant role \cite{McGuire:1981:835, Gangopadhyay:1986:2998, Starace:1987:1705, Lhuillier:1987:4747, Fink:1990:3801, Manakov:1999:3747, Makris:2008:023401, Petrov:2016:033408, Lagutin:2017:063414}. With the possibility of producing intense high-photon energy beams, the multiphoton inner-shell ionisation came in focus as well \cite{Novikov:2001:4857, Koval:2003:375, Florescu:2011:033425, Hofbrucker:2016:063412}. In these relativistic regimes, the wavefunction contraction has been found to play an important role, while contributions of higher multipoles remained to be generally less important, similarly as in the outer-shell ionisation \cite{Hofbrucker:2017:125}.

In single-photon ionisation, effects beyond the electric dipole approximation have been explored at Cooper minima \cite{Cooper:1964:762}. Such a Cooper minimum arises at a incident photon energy, where the dominant dipole element passes through a local minimum. This minimum does not only influence the shape of the total cross section, but more significantly, can strongly affect the photoelectron angular distributions \cite{White:1979:1661}. It has been shown theoretically before \cite{Johnson:1978:1167, Kim:1981:1326}, that strong anisotropic effects can be observed near Cooper minima due to relativistic and correlation effects, which are necessary to explain experimental measurements \cite{Dehmer:1976:1049, White:1979:1661, Ilchen:2018:4659}. Moreover, large nondipole contribution has been predicted in the XUV + IR two-photon above-threshold ionisation of neon $1s$ electron, when XUV photon energy matches the Cooper minimum \cite{Grum:2014:043424}.

It has been observed, that a similar Cooper minimum is also present in total cross sections of multi-photon ionisation processes \cite{Moccia:1983:2737, Saenz:1999:5629, Nikolopoulos:2006:043408}, where it appears in a form of a local minimum. In our recent work \cite{Hofbrucker:2019:011401}, we showed that for the two-photon ionisation of an $nl$ ($l>0$) electron, these \textit{nonlinear Cooper minima} can be always found between any pair of $n'(l+1)$ and $(n'+1)(l+1)$ virtual intermediate resonances, where the dominant channel $l \rightarrow l+1 \rightarrow \varepsilon (l+2)$ vanishes. Two-photon ionisation at this minimum is then described by the channel(s) $l\rightarrow l - 1 \rightarrow \epsilon l$, and in the case of ionisation by circularly polarised laser, only the electrons with orbital momentum projections $m_l<l-1$ get ionised. The photoion in this case appears to be in an aligned state, which, in the case of inner-shell ionisation, is imprinted in the polarisation degree of subsequent fluorescence light \cite{Hofbrucker:2019:011401}.
As the nonlinear Cooper minima are a property of the fundamental two-(or generally multi-)photon ionisation process itself, they are expected to strongly influence many other observables such as photoelectron \cite{Hofbrucker:2017:013409, Hofbrucker:2018:053401} or Auger electron polarisation, as well as their angular distributions.

In this paper, we consider the case of atomic $K$-shell ionisation ($l=0$). In this case, the ionisation by circularly polarised light proceeds only via the single nonrelativistic channel $s\rightarrow p \rightarrow \varepsilon d$, and therefore, all relative characteristics such as photoion polarisation and photoelectron angular distribution stay always the same. However, when we account for the spin orbit interaction, only the $s$ electron with one of the initial spin projections can be ionised via the single channel $s_{1/2} \rightarrow p_{1/2} \rightarrow \varepsilon d_{3/2}$ in the electric dipole approximation, while both electrons can be ionised over an intermediate $p_{3/2}$ state. Consequently, passing through a Cooper minimum between a pair of $n'p_{3/2}$ and $(n'+1)p_{3/2}$ resonances, one might expect a pure depletion of one of the spin projections of the initial $s$ electron. However, this conclusion holds true only in the electric dipole approximation. Here, we will show that in the case of two-photon ionisation of $K$-shell at nonlinear Cooper minima, accounting for beyond-dipole contributions becomes inevitable.
In other words, the fragile spin nature of the nonlinear Cooper minimum of a fine-structure channel gives us the opportunity to access multipole contributions in nonlinear light-matter interaction processes. 
To demonstrate the breakdown of the electron-dipole approximation on examples, we propose similar conditions as considered in the recent experiments, where either ion \cite{Sorokin:2007:213002, Richter:2010:194005, Doumy:2011:083002, Fukuzawa:2013:173005}, or fluorescence \cite{Tamasaku:2014:10.1038, Zitnik:2014:193201, Szlachetko:2016:33292, Ghimire:2016:043418, Tamasaku:2018:083901} yields were detected as a signature of two-photon $K$ shell ionisation. However, instead of solely detecting the yields, we suggest to additionally carry out measurements of degree of polarisation of photoions or fluorescence photons. 

To explain the suggested scenario in detail, we start with theoretical description of two-photon ionisation of $s-$electrons with the use of density matrix theory. We use this theoretical approach to demonstrate the breakdown of the electric dipole approximation at nonlinear Cooper minimum on examples of nonsequential two-photon ionisation of neutral germanium atoms and helium-like neons ion by right-circularly polarised light.

\section*{Methods}
In what follows, we will consistently use the typical many-electron notation in which $J$ represents the total angular momentum, $M$ its projection and $\alpha$ all further numbers that are needed for unique characterisation of an atomic state. Corresponding lower case notation is used for one-electron notation. We consider $K$-shell ionisation of an atom in an initial many-electron state $\ketm{\alpha_i J_i M_i}$ by two right-circularly polarised photons $\gamma$ with energy $\omega$. After the interaction, a photoelectron $\ketm{\bm{p}_e m_e}$, with asymptotic momentum $\bm{p}_e$ and spin projection $m_e$, is emitted from the atom, leaving it in an excited state $\ketm{\alpha_f J_f M_f}$. We can schematically represent this process as
\begin{eqnarray}
    \ketm{\alpha_i J_i M_i} +2\gamma(\omega) \rightarrow \ketm{\alpha_f J_f M_f} + \ketm{\bm{p}_e m_e}.
\end{eqnarray}
The initial atom is assumed to be in its unpolarised ground state described by the following density matrix
\begin{eqnarray}
\label{eq.InitialDensityMatrix}
\mem{\alpha_i J_i M_i}{\hat{\rho}_i}{\alpha_i J_i M_i'} = \frac{1}{[J_i]}\delta_{M_i M_i'},
\end{eqnarray}
with the notation $[J]=(2J+1)$. The two photons are assumed to come from the same source, and hence, have equivalent wavevector $\bm{k}$, as well as polarisation, which can be conveniently described in helicity ($\lambda$) representation. Since we consider right-circularly polarised photons, their polarisation state can be described by a single Stokes parameter $P_3$ denoting their degree of polarisation \cite{Blum:1981}
\begin{equation}
\label{eq.PhotonDensityMatrix}
    \mem{\bm{k} \lambda}{\hat{\rho}_{\gamma}}{\bm{k} \lambda'} = \frac{1}{2} \delta_{\lambda \lambda'}(1+\lambda P_3).
\end{equation}
The two-photon ionisation process can be analysed by measuring the properties of the emitted photoelectrons \cite{Ma:2013:164018}, or the photoions \cite{Sorokin:2007:213002, Richter:2010:194005, Doumy:2011:083002, Fukuzawa:2013:173005}. In most experiments, however, the two-photon $K$-shell ionisation of atoms has been detected by measuring subsequent fluorescence light \cite{Tamasaku:2014:10.1038, Zitnik:2014:193201, Szlachetko:2016:33292, Ghimire:2016:043418, Tamasaku:2018:083901}. We will, therefore, also examine this second process, in which the excited atom $\ketm{\alpha_f J_f M_f}$ relaxes into a lower energy state $\ketm{\alpha_0 J_0 M_0}$ by emission of fluorescent photon $\gamma_0$ with energy $\omega_{K\alpha}$
\begin{eqnarray}
    \ketm{\alpha_f J_f M_f}
    &\rightarrow& \ketm{\alpha_0 J_0 M_0} + \gamma_0(\omega_{K\alpha}).
\end{eqnarray}
In the next subsection, we will first provide a description for the simultaneous interaction of the atom with two photons. Then, we will derive the density matrices of the produced ion as well as of the fluorescence photon and analyse their properties. The results in this paper have been obtained within independent particle approximation, however, recently published many-electron code  based on multi-configuration Dirac-Fock method promises easy-to-use tools for calculations of multi-photon absorption or emission processes \cite{Fritzsche:2019:1}. Relativistic units ($\hbar=c=m=1$) are used throughout the paper, unless otherwise stated. 

\subsection*{Two-photon transition amplitude}
In second-order perturbation theory, the absorption of two photons is described by the many-electron transition amplitude
\begin{eqnarray}
\label{eq.GeneralTransitionAmplitude}
M_{J_i M_i J_f M_f m_e}^{\lambda_1 \lambda_2}(\omega)&=&~\int\kern-1.0em\sum_{\nu}\frac{
	\mem{\alpha_fJ_fM_f,\bm{p}_e m_e}
		{\hat{\mathcal{R}}(\bm{k},\lambda_2)}
		{\alpha_{\nu}J_{\nu}M_{\nu}}
	\mem{\alpha_{\nu}J_{\nu}M_{\nu}}
		{\hat{\mathcal{R}}(\bm{k},\lambda_1)}
		{\alpha_iJ_iM_i}		
		}
	{E_{i}+\omega-E_{\nu}}.
\end{eqnarray}
The one-particle transition operator $\hat{\mathcal{R}}$ from Eq. (\ref{eq.GeneralTransitionAmplitude}) can be represented in the second quantization formalism \cite{Johnson:2007}
\begin{eqnarray}
\label{eq.OperatorExpansion}
\hat{\mathcal{R}}(\bm{k},\lambda)=
	\sum_{n_1 n_2}
	\mem{n_2}{\alpha_{\mu} A^{\mu}_{\lambda}(\omega)}{n_1}
	a^{\dag}_{n_2} a_{n_1},
\end{eqnarray}
where $\ketm{n_1}$ and $\ketm{n_2}$ are single-electron states, and $a^{\dag}_{n}$, $a_{n}$ represent the corresponding creation and annihilation operators, respectively. Moreover, $\alpha_{\mu}$ is the four-vector Dirac matrices and $A^{\mu}_{\lambda}(\omega)$ is the photon wavefunction. According to the particle-hole formalism, a state with a hole in a substate $\ketm{1s_{1/2} m_a}$ has the same angular momentum properties as an electron with total angular momentum projection $-m_a$. Within the independent particle approximation, the final many-electron state can be obtained using the creation and annihilation operators as
\begin{eqnarray}
\label{eq.FinalMultiElectronWaveFunction}
\ketm{\alpha_f J_f M_f, \bm{p}_e m_e}=	\sum_{m_a M} \sprm{1/2 -m_a,J_i M}{J_f M_f} 
	(-1)^{1/2-m_a}a^{\dag}_{\bm{p}_e m_e}a_{1s_{1/2} m_a}\ketm{\alpha_i J_i M}, 
\end{eqnarray}
 with typical Clebsch-Gordan notation $\sprm{..,..}{..}$. For future analysis, it is convenient to perform the decomposition of the photon field into its electric ($p=1$) and magnetic ($p=0$) components, with a multipole order $J$ and its projection $M$
\begin{eqnarray}
\label{eq. MultipoleExpansion}
\bm{A}_{\lambda}(\omega)=
	\sqrt{2\pi} \sum_{J M p} i^{J-p}
	[J]^{1/2}(-\lambda)^p D^{J}_{M\lambda}(\bm{\hat{k}})
	\bm{a}^{(p)}_{JM}(\bm{r}),
\end{eqnarray}
 where $D_{M_1 M_2}^J (\bm{\hat{k}})$ is the Wigner D-matrix, we choose $\bm{\hat{k}}=\bm{\hat{z}}$, and hence, $D^{J}_{M\lambda}(\bm{\hat{z}})=\delta_{M \lambda }$. The term $J=p=1$ represents the electric dipole approximation, while other values of $J$ and $p$ will be refered to as contributions beyond the electric dipole approximation. Further simplification of the transition amplitude can be achieved by expanding the photoelectron wavefunction into its partial waves \cite{Eichler:1995}
\begin{equation}\label{A6}
\ketm{\bm{p}_e m_e}=\frac{1}{\sqrt{|\bm{p}_e|\varepsilon_e}}\sum_{j m_j}\sum_{l m_l} i^l e^{-i\Delta_{jl}} \sprm{l m_l, 1/2m_e}{j m_j}Y^{*}_{l m_l}(\bm{\hat{p}}_e) \ketm{\varepsilon_e j l m_j},
\end{equation}
 with a phase factor $\Delta_{jl}$, $\varepsilon_e = \sqrt{\bm{p}_e^2+m^2 }$ and spherical harmonics $Y_{l m_l}$. From the above expansions, as well as expressions (\ref{eq.OperatorExpansion}), (\ref{eq.FinalMultiElectronWaveFunction}), the many-electron amplitude (\ref{eq.GeneralTransitionAmplitude}) can be simplified to an amplitude describing one-electron transition only $M^{\lambda_1 \lambda_2}_{J_i M_i J_f M_f m_e}$ \cite{Hofbrucker:2016:063412}
\begin{eqnarray}
\label{FinalTransitionAmplitude}
M^{\lambda_1 \lambda_2}_{J_i M_i J_f M_f m_e} (\omega)&=&  
	\sum_{\substack{p_1 J_1\\p_2 J_2}} \sum_{\substack{j_n l_n m_n}}
	i^{J_1-p_1+J_2-p_2} \sqrt{\frac{[J_1, J_2]}{2[j_n]}} 
	(-\lambda_1)^{p_1} (-\lambda_2)^{p_2}
	\sum_{\substack{jm_j\\lm_l}} (-i)^l e^{i\Delta_{jl}} Y_{l m_l}(\bm{\hat{p}}_e) 
	\sprm{l m_l,1/2 m_e}{j m_j}\\\nonumber 
	&\times&  (-1)^{j-m_j} 
	\sprm{jm_j,J_2-\lambda_2}{j_nm_n}
	\sum_{m_a} \sprm{1/2 -m_a, J_i M_i}{J_f M_f} \sprm{j_n m_n, J_1-\lambda_1}{1/2 m_a}
	U^{(j_n)}_{l_j}(p_1 J_1 p_2 J_2),
\end{eqnarray}
where the matrix element describes the transition of the single active electron from the initial state $\ketm{1s_{1/2}}$ through virtual intermediate state $\ketm{n_n j_n l_n}$ to the final continuum state $\ketm{\epsilon_e  j l}$ and the reduced amplitudes $U^{(j_n)}_{l_j}(p_1 J_1 p_2 J_2)$ are given by
\begin{equation}\label{Eq.TransitionAmplitude}
    U^{(j_n)}_{l_j}(p_1 J_1 p_2 J_2)=
    \SumInt_{n_n} \frac{
        \memred{\epsilon_e j l}{\bm{\alpha}\cdot\bm{a}_{J_2}^{(p_2)}}{n_n j_n l_n}
        \memred{n_n j_n l_n}{\bm{\alpha}\cdot\bm{a}_{J_1}^{(p_1)}}{1s_{1/2}}
        }
    {E_{1s}+\omega-E_{n_n j_n}}.
\end{equation}
 In order to calculate the above transition amplitudes, we solve the Dirac equation with a screening potential, which partially accounts for the inter-electronic interaction. We use the core-Hartree potential, which corresponds to a potential created by all bound electrons except of the active electron.

There are specific incident photon energies $\omega$ between two intermediate level resonances $E_{n_n j_n}<\omega<E_{n'_n j_n}$ for which one of the amplitudes $U^{(j_n)}_{l_j}=0$, due to balance of the its spectral contributions\cite{Hofbrucker:2019:011401}. The photon energy, at which the dominant ionisation channels drops to zero, we call \textit{nonlinear Cooper minimum}, and we will demonstrate that at this energy all possible observables of the two-photon ionisation process are strongly influenced. Hence, studying the response of atoms at this energy is the key to understanding nonlinear light-matter interaction. 


\begin{figure*}
    \centering
    \includegraphics[scale=0.5]{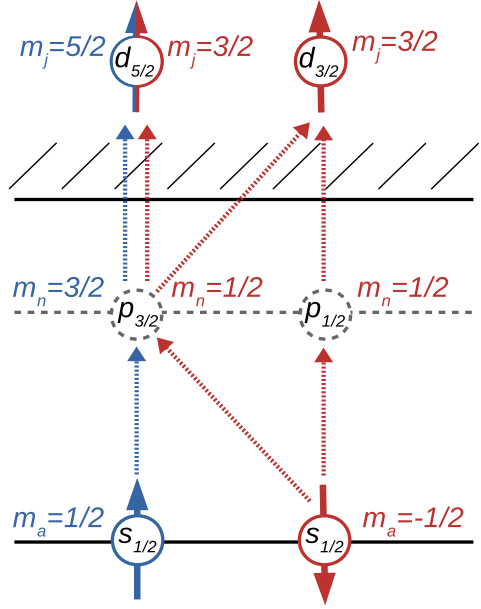}
    \caption{Possible electric dipole ionisation channels in nonsequential two-photon ionisation of an $s_{1/2}$ state by two right-circularly polarised photons. While both spin projections of the initial state can be promoted to a final $d_{5/2}$ partial wave, selection rules dictate that only spin-down electron can be ionised into a partial wave with $d_{3/2}$ symmetry. This figure has been generated using Mathematica 11.0.0.0 (\href{https://www.wolfram.com/mathematica/}{wolfram.com/mathematica}) and Inkscape 0.92 (\href{https://inkscape.org/}{inkscape.org}).}
    \label{fig:Channels}
\end{figure*}

\subsection*{Spin population of the produced ion}
Upon the simultaneous absorption of the two photons by the atom, one of the initially bound $1s$ electrons is promoted into continuum. The state of the system after ionisation $\ketm{\alpha_f J_f M_f, \bm{p}_e m_e}$ therefore consists of a singly charged ion in an excited state and a photoelectron. The density matrix of this state can be written as
\begin{eqnarray}
\label{eq.FinalDensityMatrix}
\mem{\alpha_f J_f M_f, \bm{p}_e m_e}{\hat{\rho}_f}{\alpha_f J_f M_f',  \bm{p}_e m_e'}  
	&=&\sum_{\substack{M_i \lambda_1 \lambda_2\\M_i' \lambda_1' \lambda_2'}}
		M^{\lambda_1 \lambda_2}_{J_i M_i J_f M_f m_e}(\omega)
		M^{\lambda_1' \lambda_2'*}_{J_i M_i' J_f  M_f' m_e'}(\omega)
		\mem{\alpha_i J_i M_i}{\hat{\rho}_i}{\alpha_i J_i M_i'} 
    \mem{\bm{k} \lambda_1}{\hat{\rho}_{\gamma}}{\bm{k} \lambda_1'}\nonumber\\
    &\times&
    \mem{\bm{k} \lambda_2}{\hat{\rho}_{\gamma}}{\bm{k} \lambda_2'}.
\end{eqnarray}
Using Eqs. (\ref{eq.InitialDensityMatrix}) and (\ref{eq.PhotonDensityMatrix}) as well as the fact that we do not consider any detection of the photoelectron, we trace out the photoelectron degrees of freedom from the density matrix (\ref{eq.FinalDensityMatrix}) and obtain density matrix corresponding to the excited photoion only
\begin{eqnarray}
\label{eq.SAEIonDensityMatrix}
    \mem{\alpha_f J_f M_f}{\hat{\rho}_f}{\alpha_f J_f M_f'} 
    &=& 
        \sum_{\substack{\lambda_1 \lambda_2 \\ \lambda_1' \lambda_2'}}
    \frac{1}{4[J_i]}
    \delta_{\lambda_1 \lambda_1'}\delta_{\lambda_2 \lambda_2'}
    (1+\lambda_1 P_3)(1+\lambda_2 P_3)
    \int d\Omega_e
    \sum_{M_i m_e}
    M^{\lambda_1 \lambda_2}_{J_i M_i J_f M_f m_e}(\omega)
    M^{\lambda_1' \lambda_2' *}_{J_i M_i J_f M_f' m_e}(\omega).
\end{eqnarray}
If we consider measuring the ion yield due to two-photon ionisation, we need to calculate the quantity of total two-photon ionisation cross section, which is simply given by the trace of the photoion density matrix
\begin{equation}
    \sigma (\omega) = \frac{32\pi^5\alpha^2}{\omega^2}\sum_{M_f} \mem{\alpha_f J_f M_f}{\hat{\rho}_f}{\alpha_f J_f M_f}.
\end{equation}
However, in order to detect ion polarisation, we define the reduced statistical tensor $\mathpzc{A}_{k0}$, which reflects the magnetic population of the produced ion. It is given by the statistically weighted sum of the diagonal elements of the ion density matrix
\begin{eqnarray}
\mathpzc{A}_{k 0} (J_f)&=& \frac{1}{\sum_{M_f} \mem{\alpha_f J_f M_f}{\hat{\rho}_f}{\alpha_f J_f M_f}}\sum_{M_f M_f'}(-1)^{J_f-M_f} \sprm{J_f M_f, J_f -M_f'}{k 0}\mem{\alpha_f J_f M_f}{\hat{\rho}_f}{\alpha_f J_f M_f'}.
\end{eqnarray}
To get a better feeling about the properties of this observable, let us provide the expression for ion orientation after two-photon ionisation of a $1s$ of a closed-shell atom in the electric dipole ($J_1=J_2=1$ and $p_1=p_2=1$) approximation
\begin{eqnarray}
\label{eq.IonAlignment}
\mathpzc{A}_{1 0} (1/2)&=&  \frac{6P_3 \Big[(U_{d_{3/2}})^2-4(U_{d_{5/2}})^2 \Big]}
        {6\Big[(U_{d_{3/2}})^2+6(U_{d_{5/2}})^2 \Big]+5 \delta P_3 \delta U^2 },
\end{eqnarray}
where $U_{d_{5/2}}=U^{(3/2)}_{d_{5/2}} (E1E1)$, $U_{d_{3/2}}=\big[\sqrt{10}U^{(1/2)}_{d_{3/2}}(E1E1)+U^{(3/2)}_{d_{3/2}}(E1E1)\big]$, $\delta P_3=1-(P_3)^2$ describes the lack of purity of the degree of circular polarisation of the incident light, and 
$\delta U^2 = \Big[ 
    8 (U^{(1/2)}_{s_{1/2}})^2
    -4U^{(1/2)}_{s_{1/2}} U^{(3/2)}_{s_{1/2}}
    +(5U^{(3/2)}_{s_{1/2}})^2
    -4(U^{(1/2)}_{d_{3/2}})^2
    -2\sqrt{10}U^{(1/2)}_{d_{3/2}}U^{(3/2)}_{d_{3/2}}
    +2(U^{(3/2)}_{d_{3/2}})^2
    -3(U^{(3/2)}_{d_{5/2}})^2
    \Big]$ with the $(E1E1)$ notation dropped for practical reasons.
Generally, the parameter is near zero, which can be understood from Eq. (\ref{eq.IonAlignment}) and the fact, that $U_{d_{3/2}}\approx 2 U_{d_{5/2}}$ due to angular momentum weight factors. However, it reaches $\mathpzc{A}_{1 0} (1/2)=\pm1$ whenever only one of the spin projections of the $K-$shell electrons can be ionised. This happens only if the incident beam is fully polarised ($P_3=\pm1$) and when the ionisation channel $U_{d_{3/2}}$ strongly dominates the process. One important example, where this channel dominates is at the nonlinear Cooper minimum. A detailed explanation of this phenomenon will shown together with demonstration of its effects on examples in the results section. 
%

\subsection*{Properties of subsequent fluorescence photons}
Equation (\ref{eq.SAEIonDensityMatrix}) fully describes all properties of the excited photoion. However, in the case of inner-shell ionisation, where the created hole is surrounded by electrons in higher orbitals, it is often more convenient to detect the photoion state indirectly via subsequent fluorescent emission. We, therefore, need to describe the radiative decay of the excited ion into its lower energy state $\ketm{\alpha_0 J_0 M_0}$ via $K\alpha$ emission. The photon density matrix describing the fluorescent photon with momentum $\bm{k}_0$ in the helicity representation $\lambda_0$ has the general form

\begin{equation}
    \mem{\bm{k}_0 \lambda_0}{\hat{\rho}_{\gamma_0}}{\bm{k}_0 \lambda_0'} = \sum_{\substack{M_f M_f'M_0}}
    \mem{\alpha_f J_f M_f}{\hat{\rho}_f}{\alpha_f J_f M_f'} 
    \mem{\alpha_0 J_0 M_0}{\hat{\mathcal{R}}(\bm{k}_0,\lambda_0)}{\alpha_f J_f M_f}
    \mem{\alpha_0 J_0 M_0}{\hat{\mathcal{R}}(\bm{k}_0,\lambda_0')}{\alpha_f J_f M_f'}^*.
\end{equation}
This general expression can be simplified with the use of the photon field of Eq. (\ref{eq. MultipoleExpansion}) and the Wigner-Eckart theorem \cite{Varshalovich:1988}. We obtain
\begin{eqnarray}
    \mem{\bm{k}_0 \lambda_0}{\hat{\rho}_{\gamma_0}}{\bm{k}_0 \lambda_0'} &=& 
    2\pi \sum_{\substack{J M p \\ J' M' p'}}\sum_{\substack{M_f M_f'M_0}}
    \lambda_0^p \lambda_0'^{p'} i^{J-p+J'-p'}
    D_{M \lambda_0}^J(\theta, \phi) D_{M' \lambda_0'}^{J'*}(\theta, \phi) 
    \sprm{J_f M_f, J M}{J_0 M_0}\\\nonumber
    &\times& \sprm{J_f M_f', J' M'}{J_0 M_0} 
    \mem{\alpha_f J_f M_f}{\hat{\rho}_f}{\alpha_f J_f M_f'} [J_0][J,J']^{1/2} \\\nonumber
    &\times&
    \memred{\alpha_0 J_0 }{\bm{\alpha}\cdot\bm{a}_{J}^{(p)}}{\alpha_f J_f }
    \memred{\alpha_0 J_0 }{\bm{\alpha}\cdot\bm{a}_{J'}^{(p')}}{\alpha_f J_f }^*,
\end{eqnarray}
 with the polar and azimuthal angles $\theta$ and $\phi$ with respect to the $\bm{\hat{z}}$ axis, which was chosen to be along the incoming photon direction $\bm{\hat{k}}$. In the case of two-photon ionisation of an $s$ electron by circularly polarised incident light the dependence on $\phi$ vanishes due to symmetry reasons and the degree of circular polarisation of the subsequent fluorescence photon $P^{(J_0)}_{3}$, resulting in a final ion state with total 
 angular momentum $J_0=1/2$ and $J_0=3/2$ corresponding to $K_{\alpha_2}$ and $K_{\alpha_1}$ can be written in the electric dipole approximation as 
\begin{eqnarray}
\label{eq.Stokes}
P_3^{(1/2)}(\theta)&=& \mathpzc{A}_{1 0} (1/2) \textrm{cos}\theta,\\
P_3^{(3/2)}(\theta)&=&-\frac{1}{2}\mathpzc{A}_{1 0} (1/2) \textrm{cos}\theta.
\end{eqnarray}
 The simple expressions above clearly demonstrate, that the fluorescence photon is fully determined by the orientation of the ion and carries the spin polarisation information. Since the $K\alpha_1$ and $K\alpha_2$ fluorescent photons have different energies, they are experimentally distinguishable \cite{Tamasaku:2014:10.1038}. This means that the degrees of circular polarisation corresponding to each radiative decay can be measured and analysed separately. From the above expression, it is clear that detecting the fluorescence light emitted by the decay of the $p_{1/2}$ electron will yield larger signal in measurements of the degree of polarisation than the one emitted by the decay of the $p_{3/2}$ electron.

\section*{Results}

In the following section, we study numerically the polarisation  of neutral germanium and helium-like neon following the non-sequential two-photon $K$ shell ionisation by right-circularly polarised light. Within the electric dipole approximation, there are only few possible ionisation channels which the active electron can undertake (see Fig. \ref{fig:Channels}). While electrons with both spin-up and spin-down can be ionised from the $K$ shell via the intermediate state with $p_{3/2}$ symmetry into a final $d_{5/2}$ partial wave, only spin-down electron can be ionised via the $p_{1/2}$ or $p_{3/2}$ intermediate states to the $d_{3/2}$ partial wave. Therefore, under conditions where the channel with final $d_{3/2}$ symmetry strongly dominates the process, only the spin-down electron will ionised and the produced ion will be strongly oriented. This generally happens if the incident beam energy matches either an intermediate $p_{1/2}$ level resonance, or if it is tuned to nonlinear Cooper minimum. Nonlinear Cooper minimum describes a photon energy where one of the ionisation channels vanishes due to balance of positive ($E_{1s}+\omega>E_{n_n j_n}$) and negative ($E_{1s}+\omega<E_{n_n j_n}$) denominators of the virtual intermediate states of Eqn. (\ref{Eq.TransitionAmplitude}). Due to the cancellation, the channel involving the final $d_{5/2}$ partial wave vanishes and only the spin-down electron will be ionised. Therefore, based on the prediction of the electric dipole approximation, the ion orientation parameter at nonlinear Cooper minimum (as well as at intermediate $p_{1/2}$ resonances) should be equal to unity. However, in the examples below, we show that numerical calculation carried out beyond the dipole approximation reveals breakdown of this prediction in the case of $K-$shell ionisation. The numerical results are obtained in the independent particle approximation with the core-Hartree screening potential. In order to estimate the accuracy, we also perform the calculations with the Perdew-Zunger, Kohn-Sham, and Dirac-Slater potentials~\cite{Perdew:1981:5048, Kohn:1965:A1133, Slater:1951:385}. In all the potentials, the nonlinear Cooper minima, degrees of polarisation as well as the cross sections vary not more than by 10\%.

\subsection*{Polarisation of fluorescent light after two-photon ionisation of $1s$ electron of Germanium}
\begin{figure*}
    \centering
    \includegraphics[scale=0.68]{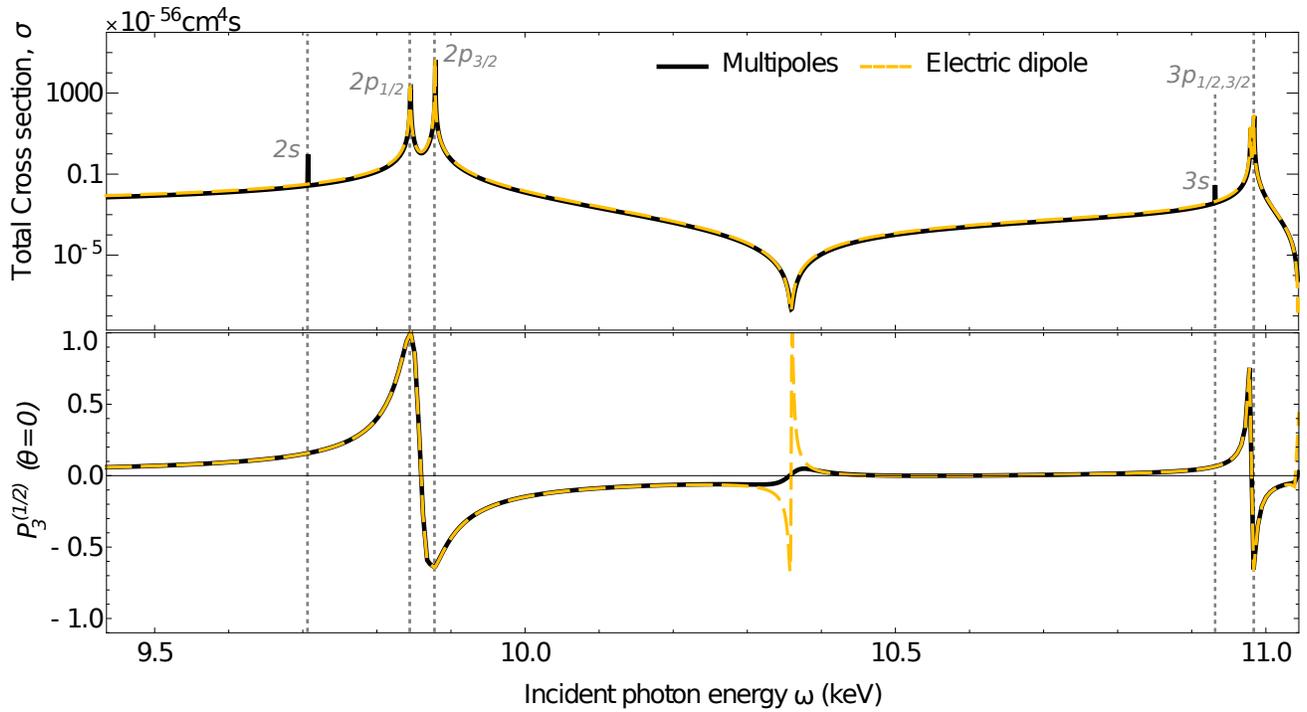}
    \caption{Direct two-photon ionisation of germanium atom by two right-circularly polarised photons within electric dipole approximation (dot-dashed yellow), and including higher multipole orders (full, black). Top: Total photoionisation cross section as function of incident photon energy. The nonlinear Cooper minimum is reflected into the cross section in a form of a local minimum around $\omega= 10.35$~keV. Bottom: Degree of circular polarisation $P^{(1/2)}_3(\theta)$ of subsequent $K\alpha_2$. A clear breakdown of the dipole approximation is visible at nonlinear Cooper minimum.}
    \label{fig:CrossSectionsP3}
\end{figure*}
\begin{figure*}
    \centering
    \includegraphics[scale=0.3]{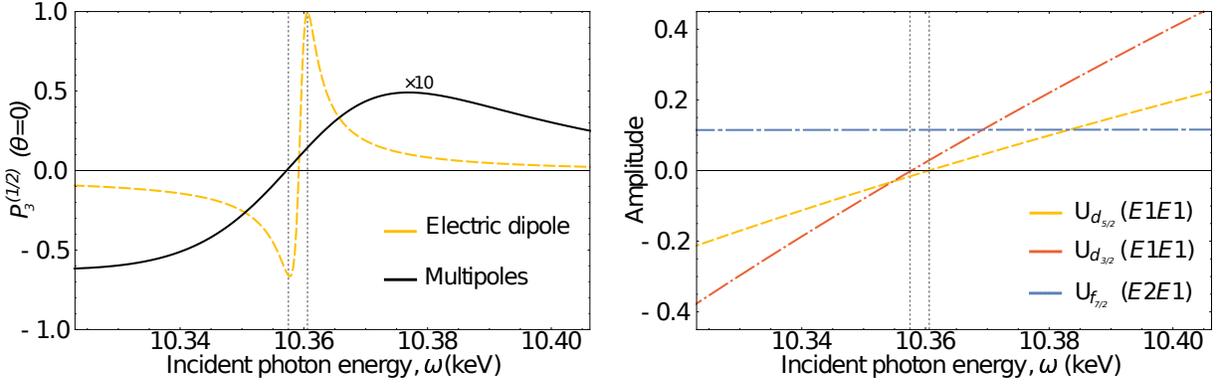}
    \caption{Left: Same as Fig. \ref{fig:CrossSectionsP3}, but zoomed into the polarisation signal at nonlinear Cooper minimum. Right: Both electric dipole transition amplitudes as well as one of the multipole amplitudes (in atomic units) as functions of incident photon energy. The peaks in the polarisation signal of the electric dipole calculation on the left part of the figure can be matched with the zero values of electric dipole amplitudes. }
    \label{fig:ZoomOnNCM}
\end{figure*}
In the following example, we consider ionisation of the $K$ shell electron of neutral germanium ($E_{1s}=11.1$~keV) by two-right-circularly polarised photons, with subsequent detection of $K\alpha$ fluorescent light. A similar scenario has been already realised with linearly polarised beam \cite{Tamasaku:2014:10.1038} at the SACLA free-electron laser. The experimentally determined total ionisation cross section for $\omega=5.6$ keV is  $\sigma_{\textrm{exp}} \approx 0.64 \times 10^{-59}$cm$^4$s. The uncertainty associated with the experimental value is not mentioned, however, we can assume that it is not better than 50\% (typical for cross section data). It is, therefore, in a reasonable agreement with our theoretical result $\sigma_{\textrm{theo}} \approx 2 \times 10^{-59}$cm$^4$s \cite{Hofbrucker:2016:063412}. In our calculations, we consider ionisation of the zero-spin isotopes of germanium (which accounts for about 92\% of the naturally occurring germanium) by two photons with energies in the range between $9.5-11$~keV. This range covers photon energies matching the $1s \rightarrow 2p$ and $1s \rightarrow 3p$ intermediate resonances, as well as a nonlinear Cooper minimum between the resonances. The corresponding results are presented in Fig. \ref{fig:CrossSectionsP3}. To guide the eye, the upper plot shows the well-known quantity of total cross section as a function of the incident photon energy, calculated within (dashed, yellow) and beyond (full, black) electric dipole approximation. The lower plot shows the degree of circular polarisation of the subsequent fluorescent photon measured along the axis of the incoming beam, with the same colour notation as in the upper plot. The level resonances in the upper plot describe the sequential ionisation process when the electron from a given shell is ionised by one photon, and simultaneously, the $1s$ electron is promoted into the hole by the second photon. The logarithmic scale of the figure also reveals the nonlinear Cooper minimum in the total cross section in a form of a local minimum, which lies around $\omega=10.35$~keV. The position of the nonlinear Cooper minimum could be, therefore, determined from measurements of the total cross section. However, it is also clear from the figure, that apart from dipole forbidden transitions at the $1s \rightarrow 2s$ and $1s \rightarrow 3s$ resonances, the calculations within the electric dipole approximation are in a very good agreement with the multipole calculations. Hence, it would be very challenging to access information about the multipole transitions from measurements of the total cross section, instead, other observables need to be inspected.

The lower plot of Fig. \ref{fig:CrossSectionsP3}, reveals hitherto elusive information. Specifically, clear deviations in the degree of polarisation appear for either for photon energies matching an intermediate $p_{1/2,3/2}$ level (sequential ionisation) resonances, or at nonlinear Cooper minimum. While the degree of circular polarisation at level resonances agrees between electric dipole and multipole calculations, a breakdown of the dipole approximation is clearly visible at the nonlinear Cooper minimum. The high degree of polarisation of the fluorescent photons at this point drops strongly due to the contributions of the generally weak multipole ionisation channels. This breakdown of the electric dipole approximation can be better seen in figure \ref{fig:ZoomOnNCM}, which shows the degree of circular polarisation only in the vicinity of the nonlinear Cooper minimum. The left subfigure shows the degree of polarisation of the fluorescence photon within the electric dipole approximation as well as with inclusion of higher multipoles. The exact position of the Cooper minimum of each channel is marked with a dashed vertical lines. The right side of the figure shows the two electric dipole and one (of many) multipole multipole channels, and demonstrates that the multipole contributions dominate the process at the nonlinear Cooper minimum. It is due to the contributions of electric quadrupole transitions that both initial electrons can be ionised, and the produced ion is no longer strongly oriented. As a consequence, the subsequent fluorescent photons also posses low degree of polarisation. 

\subsection*{Ion orientation after two-photon ionisation of Ne$^{8+}$}
\begin{figure*}
    \centering
    \includegraphics[scale=0.43]{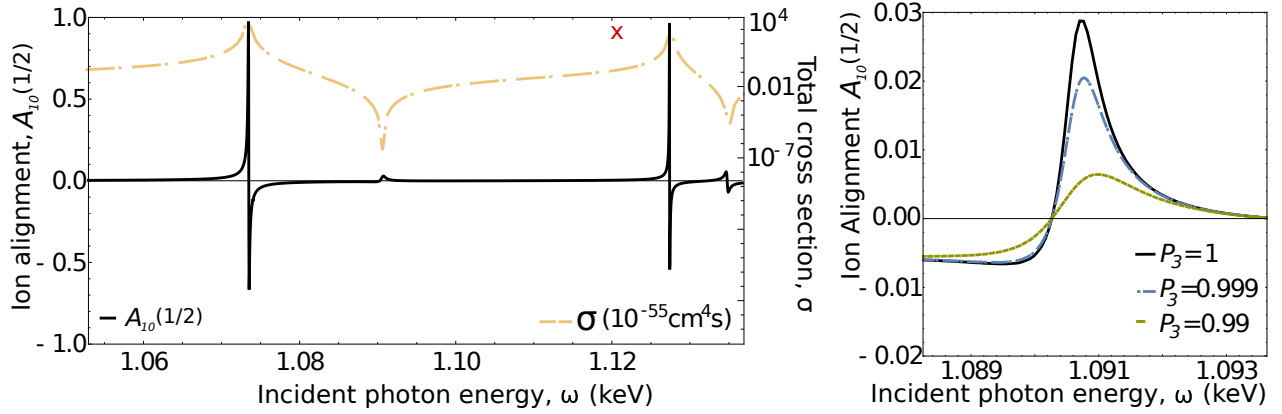}
    \caption{Ion orientation after nonsequential two-photon ionisation of Ne$^{8+}$ by two right-circularly polarised photons. Left: the total cross section as well as the orientation parameter $\mathpzc{A}_{1 0} (1/2)$ as a function of incident beam energy. The experimentally determined cross section \cite{Doumy:2011:083002} is marked with a red cross. Right: Sensitivity of the ion orientation to the purity of polarisation of the incident photons is shown in detail. }
    \label{fig:IonPolarisation}
\end{figure*}
Since, measurements of fluorescence are not always possible, other methods of measuring the ion orientation need to be carried out. The polarisation state of the ion can be detected directly with a Stern-Gerlach analyser \cite{Suzuki:2008:022902}. Here, we consider the ion detection technique to study a similar scenario as in the experiments by Doumy \textit{et al.} \cite{Doumy:2011:083002}, who reported the total cross section of two-photon ionisation of Ne$^{8+}$. The experimentally determined cross section was orders of magnitude greater than theoretically predicted values \cite{Novikov:2001:4857, Sytcheva:2012:023414, Hofbrucker:2016:063412}. We here suggest, that carrying out a similar experiment (but with a measurement of the ion orientation) could allow to test the theoretical agreement with the experiment at higher accuracy, and thus, elucidate the reason of the discrepancy. 

Figure \ref{fig:IonPolarisation} shows the total cross section as well as the ion orientation of ionisation of Ne$^{8+}$ by two right-circularly polarised photons. As seen from this figure, a nonlinear Cooper minimum appears for incident photon energy about 20~eV lower than the one used in Ref. \cite{Doumy:2011:083002}.  A measurement of this Cooper minimum with circularly polarised beam at this point, would allow one to accurately test theoretical models. The ion orientation posses the same properties as the degree of circular polarisation of fluorescence (see Eq. (\ref{eq.Stokes})). It is therefore clear, that also ion orientation will be sensitive to higher multipoles at Cooper minimum. 
Instead of reviewing the breakdown of the electric dipole approximation, in this example, we demonstrate the sensitivity of the ion orientation to the polarisation purity of the incident beam. The right side of Fig. \ref{fig:IonPolarisation} shows that de-tuning the purity of incident polarisation by mere 0.1\% results in a drop of the ion orientation at nonlinear Cooper minimum by around 30\%. If the incident beam is only 99\% circularly polarised, the ion polarisation decreases to about 20\% of the value of pure polarised beam. This high sensitivity of the ion orientation appears uniquely at the nonlinear Cooper minimum, as the influence of other channels increases quickly with polarisation de-tuning. At photon energies matching an intermediate level resonances, the process is effectively determined by one of the channels only, hence, all other channels are negligible. Due to this fact, the ion orientation is no longer extremely sensitive to the de-tuning the polarisation purity of the incident light. The purity of ion orientation at these resonant photon energies could be, however, influenced by the level widths of the fine-structure levels. 

\subsection*{Experimental consideration}
The above mentioned findings were demonstrated on examples, which were based on already performed experiments \cite{Doumy:2011:083002, Tamasaku:2014:10.1038}. In contrast to the experiments, we suggest performing the experiments with circularly polarised beams, which are already available at Linac Coherent Light Source (LCLS) at Standford \cite{Lutman:2016:468} as well as FERMI at Trieste \cite{Allaria:2013:913}. Moreover, number of other free-electron facilities include polarisation control in their upgrade plans\cite{Schneidmiller:2013:110702, Faatz:2017:1114}. With these experimental possibilities, the nonlinear Cooper minimum can play a key role in detailed understanding of nonlinear light-matter interaction. Moreover,  the degree of ion orientation (or degree of circular polarisation of fluorescent light) has been found to be extremely sensitive to the polarisation purity of ionising light and hence could be used for measuring the polarisation purity of free-electron laser beams.

\section*{Conclusion}
The concept of Cooper minimum was generalised to two-photon ionisation of inner-shell electrons, where it originates from fine-structure splitting of the $2p$ shell into the $2p_{1/2}$ and $2p_{3/2}$ subshells. The exact positions of the minima can be experimentally detected by measuring the ion, electron or fluorescent yields, and can serve as a sensitive tool to test the agreement between theory and experiment. It is worth noting that the polarisation of the incident light does not play any role in the position of the nonlinear Cooper minimum. The Cooper minima will appear in the total cross sections (and electron, photon or ion yields) for ionisation by beam of arbitrary polarisation.

It has been shown that the nonlinear Cooper minima can be utilised to access the multipole contributions from measurements of polarisation properties of the produced photoion. According to predictions based on the electric dipole approximation, it is expect that tuning the incident photon energy to the Cooper minimum in two-photon ionisation by circularly polarised light results in pure depletion of one spin projection of the initially bound $1s$ electrons, and hence, leaves the ionised atom in an oriented state. We showed that by inclusion of electric quadrupole interaction, dramatic drop of orientation purity is obtained. This breakdown of the electric dipole approximation can be readily experimentally detected either directly by Stern-Gerlach analyzer, or by means of subsequent $K\alpha-$fluorescence, which has the information about the ion orientation imprinted in its polarisation degree.


\section*{Acknowledgements}
We acknowledge the support from the Bundesministerium f\"ur Bildung und Forschung (Grant No. 05K16FJA). 

\section*{Author contributions statement}
J.H. wrote the manuscript and performed the calculations and analysis, A. V. V. conceived the original research objectives and contributed ideas in the manuscript, S. F. supervised the project throughout its duration. All authors reviewed the manuscript. 

\section*{Additional information}
\textbf{Competing interests:} The authors declare no competing interests.

\end{document}